\documentstyle[preprint,aps,epsf,psfig,eqsecnum]{revtex}
\renewcommand{\baselinestretch}{1.48}

\begin{document}
\def\be{\begin{eqnarray}}
\def\en{\end{eqnarray}}
\def\up{\uparrow}
\def\dw{\downarrow}
\def\non{\nonumber}
\def\la{\langle}
\def\ra{\rangle}
\def\ep{\varepsilon}
\def\bc{{\Lambda_b\Lambda_c}}
\def\cs{{\Lambda_c\Lambda}}
\def\bs{{\Lambda_b\Lambda}}
\def\b{\Lambda_b\to J/\psi\Lambda}
\def\j{{J/\psi}}
\def\a{{\cal A}}
\def\b{{\cal B}}
\def\bp{{\cal B'}}
\def\c{{\cal C}}
\def\cp{{\cal C'}}
\def\co{{\cal C}_1}
\def\ct{{\cal C}_2}
\def\e{{\cal E}}
\def\three{{\bf {\bar 3}}}
\def\sex{{\bf 6}}
\def\eig{{\bf 8}}
\def\ten{{\bf 10}}
\def\half{{{1\over 2}}}
\def\pr{{\sl Phys. Rev.}~}
\def\prl{{\sl Phys. Rev. Lett.}~}
\def\pl{{\sl Phys. Lett.}~}
\def\np{{\sl Nucl. Phys.}~}
\def\zp{{\sl Z. Phys.}~}

\font\el=cmbx10 scaled \magstep2
{\obeylines
\hfill IP-ASTP-06-96
\hfill May, 1997}

\vskip 1.5 cm

\centerline{\large\bf Nonleptonic Weak Decays of Bottom Baryons}
\medskip
\bigskip
\medskip
\centerline{\bf Hai-Yang Cheng}
\medskip
\centerline{ Institute of Physics, Academia Sinica}
\centerline{Taipei, Taiwan 115, Republic of China}
\bigskip
\bigskip
\bigskip
\centerline{\bf Abstract}
\bigskip
{\small   
Cabibbo-allowed two-body hadronic weak decays of bottom baryons are analyzed. 
Contrary to the charmed baryon sector, many channels of bottom baryon decays
proceed only through the external or internal $W$-emission diagrams. Moreover,
$W$-exchange is likely to be suppressed in the bottom baryon sector. 
Consequently, the factorization approach suffices to describe most of the
Cabibbo-allowed bottom baryon decays. We use the nonrelativistic quark model
to evaluate heavy-to-heavy and heavy-to-light baryon form factors at zero
recoil. When applied to the heavy quark limit, the quark model results do
satisfy all the constraints imposed by heavy quark symmetry. The decay rates
and up-down asymmetries for bottom baryons decaying into $\half^++P(V)$ and
${3\over 2}^++P(V)$ are calculated. It is found that the up-down asymmetry is 
negative except for $\Omega_b\to\half^++P(V)$ decay and for decay modes with
$\psi'$ in the final state. The prediction ${\cal B}(\Lambda_b\to 
J/\psi\Lambda)=1.6\times 10^{-4}$ for $|V_{cb}|=0.038$ is consistent 
with the recent CDF 
measurement. We also present estimates for $\Omega_c\to {3\over 2}^++P(V)$
decays and compare with various model calculations.

}

\pagebreak

\section{Introduction}
\vskip 0.15cm
While many new data of charmed baryon nonleptonic weak decays became available
in recent years, the experimental study of hadronic weak decays of bottom
baryons is just beginning to start its gear. This is best illustrated
by the decay mode $\Lambda_b\to J/\psi\Lambda$ which is interesting both
experimentally and theoretically. Its branching ratio was originally 
measured by the UA1 Collaboration to be $(1.8\pm 1.1)\times 10^{-2}$ 
\cite{UA1}. However, 
both CDF \cite{CDF93} and LEP \cite{LEP} Collaborations did not see any 
evidence for this decay. The theoretical situation is equally ambiguous:
The predicted branching ratio ranges from $10^{-3}$ to $10^{-5}$. Two early
estimates \cite{Dunietz,Cheng92} based on several different approaches
for treating the $\Lambda_b\to\Lambda$ form factors
yield a branching ratio of order $10^{-3}$. It was reconsidered in
\cite{CT96} within the nonrelativistic quark model by taking into account
the $1/m_Q$ corrections to baryonic form factors at zero recoil and
the result 
${\cal B}(\Lambda_b\to J/\psi\Lambda)=1.1\times 10^{-4}$ was
obtained (see the erratum in \cite{CT96}).
Recently, it was found that ${\cal B}(\Lambda_b\to J/\psi\Lambda)$ 
is of order $10^{-5}$ in \cite{Datta} by extracting form factors at 
zero recoil from experiment and in \cite{Guo} by generalizing the Stech's 
approach for form factors to the baryon case. This issue is finally settled
down experimentally: The decay $\Lambda_b\to J/\psi\Lambda$ is observed 
by CDF \cite{CDF96}
and the ratio of cross section times branching fraction, $\sigma_{\Lambda_b}
{\cal B}(\Lambda_b\to J/\psi\Lambda)/[\sigma_{B^0}{\cal B}(B^0\to J/\psi 
K_S)]$ is measured. The branching ratio of $\Lambda_b\to J/\psi \Lambda$ 
turns out to be $(3.7\pm 1.7\pm 0.4)\times
10^{-4}$, assuming $\sigma_{\Lambda_b}/\sigma_B=0.1/0.375$ and ${\cal B}
(B^0\to J/\psi K_S)=3.7\times 10^{-4}$. It is interesting to note that this is
also the first successful measurement of exclusive hadronic decay rate 
of bottom baryons, even though the branching ratio of 
$\Lambda_b\to\Lambda\pi$ is expected
to exceed that of $\Lambda_b\to J/\psi\Lambda$ by an order of magnitude.
Needless to say, more and more data of bottom baryon decay data will be
accumulated in the near future.

   Encouraged by the consistency between experiment and our nonrelativistic
quark model calculations for $\Lambda_b\to J/\psi\Lambda$, we would like to
present in this work a systematic study of exclusive nonleptonic decays of
bottom baryons (for earlier studies, see \cite{Rudaz,Mannel}).
Just as the meson case, all hadronic weak decays of baryons
can be expressed in terms of the following quark-diagram amplitudes 
\cite{CCT}: $\a$, the external $W$-emission diagram; $\b$, the internal 
$W$-emission diagram; $\c$, the $W$-exchange diagram and $\e$, the 
horizontal $W$-loop diagram. The external and 
internal $W$-emission diagrams are sometimes referred to as color-allowed and
color-suppressed factorizable contributions. However, baryons being made out
of three quarks, in contrast to two quarks for mesons, bring along several
essential complications. First of all, the factorization approximation that 
the hadronic matrix element is factorized into the product of two matrix
elements of single currents and that the nonfactorizable term such as 
the $W$-exchange contribution is negligible relative to the factorizable 
one is known empirically to be working reasonably well for describing the
nonleptonic weak decays of heavy mesons \cite{Cheng89}. However, this
approximation is {\it a priori} not directly applicable to the charmed 
baryon case as $W$-exchange there, manifested as pole diagrams, is no 
longer subject to helicity and color suppression.\footnote{This is 
different from the naive color suppression of internal
$W$-emission. It is known in the heavy meson case that nonfactorizable
contributions will render the color suppression of internal $W$-emission 
ineffective. However, the
$W$-exchange in baryon decays is not subject to color suppression 
even in the absence of nonfactorizable terms. A simple way to see this is
to consider the large-$N_c$ limit. Although the $W$-exchange diagram is down
by a factor of $1/N_c$ relative to the external $W$-emission one, it is
compensated by the fact that the baryon contains $N_c$ quarks in the limit 
of large $N_c$, thus allowing $N_c$ different possibilities for $W$ exchange
between heavy and light quarks \cite{Korner}.}
That is, the pole contribution can be as important as the factorizable one. 
The experimental measurement of the decay modes $\Lambda_c^+\to\Sigma^0\pi^+,~
\Sigma^+\pi^0$ and $\Lambda^+_c\to\Xi^0K^+$, which do not receive any
factorizable contributions, indicates that $W$-exchange
indeed plays an essential role in charmed baryon decays. Second, 
there are more possibilities in drawing the $\b$ and $\c$ types of 
amplitudes \cite{CCT}; in general there exist two distinct internal 
$W$-emissions and
several different $W$-exchange diagrams and only one of the internal
$W$-emission amplitudes is factorizable.

   The nonfactorizable pole contributions to hadronic weak decays 
of charmed baryons have been studied in the literature \cite{CT92,CT93,XK92}. 
In general, nonfactorizable $s$- and $p$-wave amplitudes for 
${1\over 2}^+\to {1\over 2}^++P(V)$ decays ($P$: pseudoscalar meson, $V$: 
vector meson),
for example, are dominated by ${1\over 2}^-$ low-lying baryon resonances
and ${1\over 2}^+$ ground-state baryon poles, respectively. However, the
estimation of pole amplitudes is a
difficult and nontrivial task since it involves weak baryon matrix elements
and strong coupling constants of ${1\over 2}^+$ and ${1\over 2}^-$ baryon
states. This is the case in particular for $s$-wave terms as
we know very little about the ${1\over 2}^-$ states.
As a consequence, the evaluation of pole diagrams is far more uncertain
than the factorizable terms. Nevertheless, the bottom baryon system
has some advantages over the charmed baryon one. First, $W$-exchange is
expected to be less important in the nonleptonic decays of the former.
The argument goes as follows. The $W$-exchange contribution to the total
decay width of the heavy baryon relative to the spectator diagram is of
order $R=32\pi^2|\psi_{Qq}(0)|^2/m^3_Q$ \cite{Cheng92}, where the square
of the wave function $|\psi_{Qq}(0)|^2$ determines the probability of
finding a light quark $q$ at the location of the heavy quark $Q$. Since
$|\psi_{cq}(0)|^2\sim |\psi_{bq}(0)|^2\sim (1-2)\times 10^{-2}\,{\rm GeV}^2$
\cite{Cheng92},
it is clear that $R$ is of order unity in the charmed baryon case, while it
is largely suppressed in bottom baryon decays. Therefore, although 
$W$-exchange plays a dramatic role in charmed baryon case (it even dominates
over the spectator contribution in hadronic decays of $\Lambda_c^+$ and
$\Xi_c^0$ \cite{Cheng92}), it becomes negligible in inclusive hadronic decays
of bottom baryons. It is thus reasonable to assume that the same suppression 
is also inherited in the two-body nonleptonic weak decays of bottom
baryons. Second, for charmed baryon decays, there are only a few decay modes
which proceed through external or internal $W$-emission diagram, namely,
Cabibbo-allowed $\Omega_c^0\to\Omega^-\pi^+(\rho^+),~\Xi^{*0}\bar{K}^0
(\bar{K}^{*0})$ and Cabibbo-suppressed $\Lambda_c^+\to p\phi$, 
$\Omega_c^0\to\Xi^-\pi^+(\rho^+)$. However, even at the Cabibbo-allowed
level, there already exist a significant number of bottom baryon decays 
which receive
contributions only from factorizable diagrams (see Tables II and III below)
and $\Lambda_b\to J/\psi \Lambda$ is one of the most noticeable examples.
For these decay modes we can make a reliable estimate based on the 
factorization approach as they do not involve troublesome nonfactorizable
pole terms. Moreover, with the aforementioned suppression of 
$W$-exchange, many decay channels are dominated by 
external or internal $W$-emission. Consequently, contrary to the charmed
baryon case, it suffices to apply the factorization hypothesis to describe 
most of Cabibbo-allowed two-body nonleptonic decays of bottom baryons, and
this makes the study of bottom baryon decays considerably simpler than that
in charmed baryon decays.

   Under the factorization approximation, the baryon decay amplitude is
governed by a decay constant and form factors. In order to study 
heavy-to-heavy and heavy-to-light baryon form factors, we will follow
\cite{CT96} to employ the nonrelativistic quark model to evaluate the
form factors at zero recoil. Of course, the quark model results should be
in agreement with the predictions of the heavy quark effective theory (HQET)
for antitriplet-to-antitriplet
heavy baryon form factors to the first order in $1/m_Q$ and for
sextet-to-sextet ones to the zeroth order in $1/m_Q$. The
quark model, however, has the merit that it is applicable to heavy-to-light
baryonic transitions as well and accounts for $1/m_Q$ effects for 
sextet-to-sextet heavy baryon transition.
In this paper, we will generalize the work of \cite{CT96}
to ${1\over 2}^+-{3\over 2}^+$ transitions in order to study the decays
${1\over 2}^+\to{3\over 2}^++P(V)$. As the conventional practice, we then 
make the pole dominance assumption for the $q^2$ dependence to
extrapolate the form factor from zero recoil to the desired $q^2$ point.

 The layout of the present paper is organized as follows. In Sec.~II we first 
discuss the quark-diagram amplitudes for Cabibbo-allowed bottom baryon
decays. Then with the form factors calculated using the nonrelativistic quark
model, the external and internal $W$-emission amplitudes are computed
under the factorization approximation. Results of model calculations and 
their physical implications are discussed in Sec.~III. A detail of
the quark model evaluation of form factors is presented in Appendix A and the
kinematics for nonleptonic decays of baryons is summarized in Appendix B.

\section{Nonleptonic Weak Decays of Bottom Baryons}
\subsection{Quark Diagram Classification}
  The light quarks of the bottom baryons belong to either a $\three$ or
a $\sex$ representation of the flavor SU(3). The $\Lambda_b^+$,
$\Xi_b^{0A}$, and $\Xi_b^{-A}$ form a
 $\three$ representation and they all decay weakly.  The $\Omega_b^-$,
$\Xi_b^{0S}$, $\Xi_b^{-S}$, $\Sigma_b^{+,0,-}$ form a $\sex$ 
representation; among them, however, only  $\Omega_b^-$  decays weakly.

Denoting the bottom baryon, charmed baryon, octet baryon, decuplet baryon
and octet meson by $B_b,~B_c,~B(\eig),~B(\ten)$ and $M(\eig)$, respectively,
the two-body nonleptonic decays of bottom baryon can be classified into:
\be
& (a) &~~~~ B_b(\three)\to B_c(\three)+M(\eig),  \non \\
& (b) &~~~~ B_b(\three)\to B_c(\sex)+M(\eig),  \\
& (c) &~~~~ B_b(\three)\to B(\eig)+M(\eig),   \non \\
& (d) &~~~~ B_b(\three)\to B(\ten)+M(\eig),  \non 
\en
and
\be
& (e) &~~~~ B_b(\sex)\to B_c(\sex)+M(\eig), \non \\
& (f) &~~~~ B_b(\sex)\to B_c^*(\sex)+M(\eig), \non \\
& (g) &~~~~ B_b(\sex)\to B_c(\three)+M(\eig),  \\
& (h) &~~~~ B_b(\sex)\to B(\eig)+M(\eig),  \non \\
& (i) &~~~~ B_b(\sex)\to B(\ten)+M(\eig), \non 
\en
where $B_c^*$ designates a spin-${3\over 2}$ sextet charmed baryon. In
\cite{CCT} we have given a general formulation of the quark-diagram
scheme for the nonleptonic weak decays of charmed baryons, which can 
be generalized directly to the bottom baryon case. 
The general quark diagrams for decays in (2.1) and (2.2) are: 
the external $W$-emission $\a$, internal $W$-emission diagrams $\b$ and $\bp$, 
$W$-exchange diagrams $\co,~\ct$ and $\cp$, and the horizontal $W$-loop 
diagrams $\e$ and $\e'$ (see Fig.~2 of \cite{CCT} for notation and for 
details).\footnote{The quark diagram amplitudes $\a,~\b,~\bp\cdots$ etc. in
each type of hadronic decays are
in general not the same. For octet baryons in the final state, each of the
$W$-exchange and $W$-loop amplitudes has two more independent types: the
symmetric and the antisymmetric, for example, $\c_{1A}$, $\c_{1S}$, 
$\e_A,~\e_S,\cdots$ etc. \cite{CCT}.}
The quark coming from the bottom quark decay in diagram $\bp$ contributes
to the final meson formation, whereas it contributes to the final
baryon formation in diagram $\b$. Consequently, diagram $\bp$ contains 
factorizable contributions
but $\b$ is not. Note that, contrary to the charmed baryon case,
the horizontal $W$-loop diagrams (or the so-called penguin diagrams under
one-gluon-exchange approximation) can contribute to some of 
Cabibbo-allowed decays of bottom baryons. Since the two spectator light
quarks in the heavy baryon are antisymmetrized in $B_Q(\three)$ and 
symmetrized in $B_Q(\sex)$
and since the wave function of $B(\ten)$ is totally symmetric, it is clear
that factorizable amplitudes $\a$ and $\bp$ cannot contribute to the decays 
of types (b), (d) and (g). For example, decays of type (d)
receive contributions only from the $W$-exchange and $W$-loop diagrams,
namely ${\cal C}_{2S},~\c'_S$ and $\e_S$ (see Fig.~1 of \cite{CCT}). 
There are only a few Cabibbo-allowed $B_b(\three)\to B(\ten)+M(\eig)$ decays:
\be
\Lambda_b^0\to\,D^0\Delta^0,~D^{*0}\Delta^0;~~~~\Xi_b^{0,-}\to D^0\Sigma
^{*0,-},~D^{*0}\Sigma^{*0,-}.
\en
They all only receive contributions from the $W$-exchange diagram $\c'_S$.
We have shown In Tables II and III the quark diagram amplitudes for those 
Cabibbo-allowed bottom baryon decays that do receive contributions from the
external $W$-emission $\a$ or internal $W$-emission $\bp$.

\subsection{Factorizable Contributions}
   At the quark level, the hadronic decays of bottom baryons proceed the
above-mentioned various quark diagrams. At the hadronic level, the decay
amplitudes are conventionally evaluated using factorization approximation
for quark diagrams $\a$ and $\bp$ and pole approximation for the remaining 
diagrams $\b,~\co,~\ct,\cdots$ \cite{Korner,CT92,CT93,XK92}. Among all 
possible pole contributions, including resonances and continuum states, one 
usually focuses on the most important poles such as the low-lying ${1\over 
2}^+,{1\over 2}^-$ states. However, it is difficult to make a reliable 
estimate of pole contributions since they involve baryon matrix elements and
strong coupling constants of the pole states. Fortunately, among the 32 
decay modes of Cabibbo-allowed decays ${1\over 2}^+\to{1\over 2}^++P(V)$
listed in Table II and 8 channels of ${1\over 2}^+\to {3\over 2}^++P(V)$ in
Table III, 20 of them receive
contributions only from factorizable terms.
Furthermore, as discussed in the Introduction, the $W$-exchange
contribution to the inclusive decay rate of bottom baryons relative to 
the spectator decay is of order $32\pi^2|\psi_{bq}(0)|^2/m^3_b\sim (3-5)\%$.
It is thus reasonable to assume that the same suppression persists at the 
exclusive two-body decay level. The penguin contributions $\e$ and $\e'$ to
the Cabibbo-allowed decay modes e.g., 
$\Lambda_b\to D_s^{(*)}\Lambda_c,~\Xi_b\to D_s^{(*)}
\Xi_c,~\Omega_b\to D_s^{(*)}\Omega_c$ (see Table II) can be safely neglected
since the Wilson coefficient $c_6(m_b)$ of the penguin operator $O_6$ is
of order 0.04 \cite{Buras} and there is no chiral enhancement in the
hadronic matrix element of $O_6$ due to the absence of a light meson in
the final state. Therefore, by neglecting the $W$-exchange contribution as
a first order approximation, we can make sensible predictions for most of
decay modes exhibited in Tables II and III. As for the nonfactorizable 
internal $W$-emission $\b$, there is no reason to argue that it is negligible.

   To proceed we first consider the Cabibbo-allowed decays $B_b(
{1\over 2}^+)\to B({1\over 2}^+)+P(V)$. The general amplitudes are
\be
{\cal M}[B_i(1/2^+)\to B_f(1/2^+)+P] &=& i\bar{u}_f(p_f)(A+B
\gamma_5)u_i(p_i),   \\
{\cal M}[B_i(1/2^+)\to B_f(1/2^+)+V] &=& \bar{u}_f(p_f)\ep^{*\mu}[A_1
\gamma_\mu\gamma_5+A_2(p_f)_\mu\gamma_5+B_1\gamma_\mu+B_2(p_f)_\mu] u_i(p_i),
\non
\en
where $\ep_\mu$ is the polarization vector of the vector meson. The 
QCD-corrected weak Hamiltonian responsible for Cabibbo-allowed 
hadronic decays of bottom baryons read
\be
{\cal H}_W=\,{G_F\over\sqrt{2}}\,V_{cb}V_{ud}^*(c_1O_1+c_2O_2)+(u\to 
c,~d\to s),
\en
with $O_1=(\bar{u}s)(\bar{b}c)$ and $O_2=(\bar{u}c)(\bar{b}s)$, where
$(\bar{q}_1q_2)\equiv\bar{q}_1\gamma_\mu(1-\gamma_5)q_2$. Under factorization
approximations, the external or internal $W$-emission contributions to the
decay amplitudes are given by
\be
A &=& \lambda a_{1,2}f_P(m_i-m_f)f_1(m^2_P),   \non \\
B &=& \lambda a_{1,2}f_P(m_i+m_f)g_1(m^2_P),   
\en
and
\be
A_1 &=& -\lambda a_{1,2}f_Vm_V[g_1(m_V^2)+g_2(m^2_V)(m_i-m_f)],   \non \\
A_2 &=& -2\lambda a_{1,2}f_Vm_Vg_2(m^2_V),   \non \\
B_1 &=& \lambda a_{1,2}f_Vm_V[f_1(m_V^2)-f_2(m^2_V)(m_i+m_f)],   \\
B_2 &=& 2\lambda a_{1,2}f_Vm_Vf_2(m^2_V),  \non
\en
where $\lambda=G_F V_{cb}V_{ud}^*/\sqrt{2}$ or $G_F V_{cb}V_{cs}^*/\sqrt{2}$,
depending on the final meson state under consideration, $f_i$ and $g_i$ 
are the form factors defined by ($q=p_i-p_f$)
\be
\la B_f(p_f)|V_\mu-A_\mu|B_i(p_i)\ra &=& \bar{u}_f(p_f)
[f_1(q^2)\gamma_\mu+if_2(q^2)\sigma_{\mu\nu}q^\nu+f_3(q^2)q_\mu   \non \\
&& -(g_1(q^2)\gamma_\mu+ig_2(q^2)\sigma_{\mu\nu}q^\nu+g_3(q^2)q_\mu)\gamma_5]
u_i(p_i),
\en
$m_i~(m_f)$ is the mass of the initial (final) baryon, 
$f_P$ and $f_V$ are the decay constants
of pseudoscalar and vector mesons, respectively, defined by
\be
\la 0|A_\mu|P\ra =-\la P|A_\mu|0\ra=\,if_P q_\mu,~~~~\la 0|V_\mu|V\ra=\,\la 
V|V_\mu|0\ra=\,f_Vm_V\ep^*_\mu,
\en
with the normalization $f_\pi=132$ MeV. 

   Since in this paper we rely heavily on the factorization approximation to
describe bottom baryon decay, we digress for a moment to discuss its content.
In the naive factorization approach,
the coefficients $a_1$ for the external $W$-emission amplitude and $a_2$
for internal $W$-emission are given by $(c_1+{c_2\over 3})$ and
$(c_2+{c_1\over 3})$, respectively. However, we have learned from charm 
decay that
the naive factorization approach never works for the decay rate of 
color-suppressed decay modes, though it usually operates for color-allowed
decays. For example, the predicted rate of $\Lambda_c^+\to p\phi$ in the 
naive approach is too small when compared with experiment \cite{CT92}. 
This implies that the inclusion of
nonfactorizable contributions is inevitable and necessary. If nonfactorizable
effects amount to a redefinition of the effective parameters $a_1$,
$a_2$ and are universal (i.e., channel-independent) in charm or bottom
decays, then we still have a new factorization scheme with the
universal parameters $a_1,~a_2$ to be determined from experiment. Throughout
this paper, we will thus treat $a_1$ and $a_2$ as free effective parameters.
The factorization hypothesis implies iuniversal and channle-independent 
$a_1^{\rm eff}$ and $a_2^{\rm eff}$ in charm or bottom decay.\footnote{
For $D(B)\to PP$ or $VP$ decays ($P$ denotes a pseudoscalar meson, V a 
vector meson),
nonfactorizable effects can always be lumped into the effective parameters
$a_1$ and $a_2$. For $D(B)\to VV$ and heavy baryon decays, universal
nonfactorizable terms are assumed under the factorization approximation.
The first systematical study of heavy meson decays within the framework
of improved factorization was carried out by Bauer, Stech and Wirbel
\cite{BSW}. Theoretically, nonfactorizable terms come mainly from 
color-octet currents. Phenomenological analyses of $D$ and $B$ decay data
\cite{Cheng,Kamal} indicate that while the factorization hypothesis in
general works reasonably well, the effective parameters $a_{1,2}$ do 
show some variations from channel to channel.}

   Since we shall consider heavy-to-heavy and heavy-to-light baryonic
transitions, it is clear that HQET is not adequate for our purposes: the
predictive power of HQET for baryon form factors at order $1/m_Q$ is limited 
only to antitriplet-to-antitriplet heavy baryonic transition. Hence, we will
follow \cite{CT96} to apply the nonrelativistic quark model to evaluate 
the weak current-induced baryon 
form factors at zero recoil in the rest frame of the heavy parent baryon,
where the quark model is most trustworthy. This quark model approach has 
the merit that it is applicable to heavy-to-heavy and heavy-to-light 
baryonic transitions at maximum $q^2$ and that it becomes meaningful
to consider $1/m_q$ corrections so long as the recoil momentum is smaller
than the $m_q$ scale. 

    The complete quark model results for form factors $f_i$ and $g_i$ at zero 
recoil read \cite{CT96}
\be
f_1(q^2_m)/N_{fi} &=& 1-{\Delta m\over 2m_i}+{\Delta m\over 
4m_im_q}\left(1-{\bar{\Lambda}\over 2m_f}\right)(m_i+m_f-\eta\Delta m)\non \\
&& -{\Delta m\over 8m_im_f}\,{\bar{\Lambda}\over m_Q}(m_i+m_f+\eta\Delta m),
\non  \\
f_2(q^2_m)/N_{fi} &=& {1\over 2m_i}+{1\over 4m_im_q}\left(1-{\bar{\Lambda}
\over 2m_f}\right)[\Delta m-(m_i+m_f)\eta]  \non \\
&& -{\bar{\Lambda}\over 8m_im_fm_Q}[\Delta m+(m_i+m_f)\eta],  \non \\
f_3(q^2_m)/N_{fi} &=& {1\over 2m_i}-{1\over 4m_im_q}\left(1-{\bar{\Lambda}
\over 2m_f}\right)(m_i+m_f-\eta\Delta m)   \non \\
&& +{\bar{\Lambda}\over 8m_im_fm_Q}(m_i+m_f+\eta\Delta m),  \\
g_1(q^2_m)/N_{fi} &=& \eta+{\Delta m\bar{\Lambda}\over 4}\left({1\over m_i 
m_q}-{1\over m_fm_Q}\right)\eta,  \non \\
g_2(q^2_m)/N_{fi} &=& -{\bar{\Lambda}\over 4}\left({1\over m_i m_q}-
{1\over m_fm_Q}\right)\eta,  \non \\
g_3(q^2_m)/N_{fi} &=& -{\bar{\Lambda}\over 4}\left({1\over m_i m_q}+
{1\over m_fm_Q}\right)\eta,   \non
\en
where $\bar{\Lambda}=m_f-m_q$, $\Delta m=m_i-m_f$, $q_m^2=\Delta m^2$,
 $\eta=1$ for the $\three$ baryon $B_i$, and $\eta=-{1\over 3}$ for the 
$\sex$ baryon $B_i$, and $N_{fi}$ is a flavor factor:
\be
N_{fi}=\,_{\rm flavor-spin}\la B_f|b_q^\dagger b_Q|B_i\ra_{\rm flavor-spin}
\en
for the heavy quark $Q$ in the parent baryon $B_i$ transiting into the
quark $q$ (being a heavy quark $Q'$ or a light quark) in the daughter baryon
$B_f$. It has been shown in \cite{CT96} that the quark model predictions
agree with HQET for antitriplet-to-antitriplet (e.g., $\Lambda_b\to\Lambda_c,~
\Xi_b\to\Xi_c$) form factors to order $1/m_Q$. For sextet $\Sigma_b\to
\Sigma_c$ and $\Omega_b\to\Omega_c$ transitions, the HQET predicts that 
to the zeroth order in $1/m_Q$ (see e.g., \cite{Yan})
\be
\la B_f(v',s')|V_\mu|B_i(v,s)\ra &=& -{1\over 3}\bar{u}_f(v',s')\Big\{\left[
\omega\gamma_\mu-2(v+v')_\mu\right]\xi_1(\omega)  \non \\
&& +\left[(1-\omega^2)\gamma_\mu-2(1-\omega)(v+v')_\mu\right]\xi_2(\omega)
\Big\}u_i(v,s),    \non \\
\la B_f(v',s')|A_\mu|B_i(v,s)\ra &=& {1\over 3}\bar{u}_f(v',s')\Big\{\left[
\omega\gamma_\mu+2(v-v')_\mu\right]\xi_1(\omega)   \\
&& +\left[(1-\omega^2)\gamma_\mu-2(1+\omega)(v-v')_\mu\right]\xi_2(\omega)
\Big\}u_i(v,s),    \non
\en
where $\omega\equiv v\cdot v'$, $\xi_1$ and $\xi_2$ are two universal baryon
Isgur-Wise functions with the normalization of $\xi_1$ known to be 
$\xi_1(1)=1$. From Eq.(2.12) we obtain
\be
f_1 &=& F_1+{1\over 2}(m_i+m_f)\left({F_2\over m_i}+{F_3\over m_f}\right),
\non \\
f_2 &=& {1\over 2}\left({F_2\over m_i}+{F_3\over m_f}\right), \non \\
f_3 &=& {1\over 2}\left({F_2\over m_i}-{F_3\over m_f}\right),  \\
g_1 &=& G_1-{1\over 2}(m_i-m_f)\left({G_2\over m_i}+{G_3\over m_f}\right), 
\non \\
g_2 &=& {1\over 2}\left({G_2\over m_i}+{G_3\over m_f}\right), \non \\
g_3 &=& {1\over 2}\left({G_2\over m_i}-{G_3\over m_f}\right), \non 
\en
with
\be
&& F_1=-G_1=-{1\over 3}\,[\omega\xi_1+(1-\omega^2)\xi_2],   \non \\
&& F_2=F_3={2\over 3}\,[\xi_1+(1-\omega)\xi_2],   \\
&& G_2=-G_3={2\over 3}\,[\xi_1-(1+\omega)\xi_2].  \non
\en
Since $N_{fi}=1$ and $\eta=1$ for sextet-to-sextet transition, it
follows from (2.10) that
\be
&& f_1(q_m^2) = -{1\over 3}\left[ 1-(m_i+m_f)\left({1\over m_i}+{1\over m_f}
\right)\right],   \non \\
&& f_2(q^2_m) =\,{1\over 3}\left({1\over m_i}+{1\over m_f}\right),~~~~
f_3(q^2_m) =\,{1\over 3}\left({1\over m_i}-{1\over m_f}\right),    \\
&& g_1(q^2_m)=-{1\over 3},~~~~g_2(q^2_m)=g_3(q_m^2)=0.   \non
\en
It is easily seen that at zero recoil $\omega=1$, the quark model results
(2.15) are in accord with the HQET predictions (2.13) provided that
\be
\xi_2(1)=\,{1\over 2}\,\xi_1(1)=\,{1\over 2}.
\en
This is precisely the prediction of large-$N_c$ QCD \cite{Chow}.

    Three remarks are in order. First, there are two different quark model
calculations of baryon form factors \cite{Marcial,Sing} prior to the work 
\cite{CT96}. An obvious criterion for testing the reliability of quark 
model calculations is that model results must satisfy all the constraints 
imposed by heavy quark symmetry. In the heavy quark limit, normalizations of 
heavy-to-heavy form factors and hence some relations between form factors at 
zero recoil are fixed by heavy quark symmetry. These constraints are not 
respected in \cite{Marcial}. While this discrepancy is improved in the work 
of \cite{Sing}, its prediction for $\Lambda_b\to\Lambda_c$ (or 
$\Xi_b\to\Xi_c$) form factors at order $1/m_Q$ is still too large by a 
factor of 2 when compared with HQET \cite{CT96}. Second, the flavor factor 
$N_{fi}$ (2.11) for heavy-to-light transition is usually smaller than 
unity (see Table I) due to the fact that SU(N) flavor symmetry is badly 
broken. As stressed in \cite{Sing,Kroll,Korner94}, it is important to take 
into account this flavor-suppression factor when evaluating the heavy-to-light
baryon form factors. Third, in deriving the baryon matrix elements at zero
recoil in the rest frame of the parent baryon, we have neglected the
kinetic energy (k.e.) of the quark participating weak transition relative
to its constituent mass $M_q$. This is justified in the nonrelativistic
constituent quark model even when the final baryon
is a hyperon or a nucleon. The kinetic energy of the QCD current quark
inside the nucleon at rest is of order a few hundred MeV. In the
nonrelativistic quark model this kinetic energy is essentially absorbed in
the constituent mass of the constituent quark. As a result, it is a good
approximation to neglect (k.e./$M_q$) for the constituent quarks inside
the nucleon (or hyperon) at rest. Of course, this approximation works best
for $Q\to Q'$ transition, and fairly good for $Q\to s$ or $Q\to u(d)$
transition.

   We next turn to the Cabibbo-allowed decays $B_b({1\over 2}^+)\to 
B^*({3\over 2}^+)+P(V)$ with the general amplitudes:
\be
{\cal M}[B_i(1/2^+)\to B_f^*(3/2^+)+P] &=& iq_\mu\bar{u}^\mu_f(p_f)(C+D
\gamma_5)u_i(p_i),   \non \\
{\cal M}[B_i(1/2^+)\to B_f^*(3/2^+)+V] &=& \bar{u}_f^\nu(p_f)\ep^{*\mu}[
g_{\nu\mu}(C_1+D_1\gamma_5)   \\
&& +p_{1\nu}\gamma_\mu(C_2+D_2\gamma_5)+p_{1\nu}
p_{2\mu}(C_3+D_3\gamma_5)]u_i(p_i),   \non
\en
with $u^\mu$ being the Rarita-Schwinger vector spinor for a spin-${3\over 2}$
particle. The external and internal $W$-emission contributions under 
factorization approximation become
\be
C &=& -\lambda a_{1,2}f_P[\bar g_1(m^2_P)+(m_i-m_f)\bar g_2(m_P^2)+(m_iE_f-
m^2_f)\bar g_3(m_P^2)], 
\non \\
D &=& \lambda a_{1,2}f_P[\bar f_1(m_P^2)-(m_i+m_f)\bar f_2(m_P^2)+(m_iE_f-
m^2_f)\bar f_3(m_P^2)], 
\en
and 
\be
C_i =- \lambda a_{1,2}f_Vm_V\bar g_i(m_V^2),~~~~  
D_i =\, \lambda a_{1,2}f_Vm_V\bar f_i(m_V^2),  
\en
where $i=1,2,3$, and the form factors $\bar f_i$ as well as $\bar g_i$ are 
defined by
\be
\la B_f^*(p_f)|V_\mu-A_\mu|B_i(p_i)\ra &=& \bar{u}_f^\nu
[(\bar f_1(q^2)g_{\nu\mu}+\bar f_2(q^2)p_{1\nu}\gamma_\mu+\bar f_3(q^2)
p_{1\nu}p_{2\mu})\gamma_5   \non \\ 
&& -(\bar g_1(q^2)g_{\nu\mu}+\bar g_2(q^2)p_{1\nu}\gamma_\mu+
\bar g_3(q^2)p_{1\nu}p_{2\mu})]u_i.
\en
In deriving Eq.~(2.18) we have applied the constraint $p_\nu u^\nu(p)=0$.
As before, form factors are evaluated at zero recoil
using the nonrelativistic quark model and the results are (see 
Appendix A for detail):
\be
\bar f_1(q_m^2)/N_{fi} &=& {2\over\sqrt{3}}\left(1+{\bar{\Lambda}\over 
2m_q}+{\bar{\Lambda}\over 2m_Q}\right),   \non \\
\bar f_2(q_m^2)/N_{fi} &=& {1\over\sqrt{3}m_i}\left(1+{\bar{\Lambda}\over 
2m_q}+{\bar{\Lambda}\over 2m_Q}\right),   \non \\
\bar f_3(q^2_m)/N_{fi} &=& -{1\over\sqrt{3}m_im_f}\left(1+{\bar{\Lambda}\over 
2m_q}+{\bar{\Lambda}\over 2m_Q}\right),   \non \\
\bar g_1(q^2_m)/N_{fi} &=& -{2\over \sqrt{3}},   \\
\bar g_2(q^2_m)/N_{fi} &=& -{1\over \sqrt{3}}\,{\bar \Lambda\over m_qm_i},  
\non \\
\bar g_3(q^2_m)/N_{fi} &=& -\bar f_3(q^2_m)/N_{fi}.  \non 
\en
The above form factors are applicable to heavy-to-heavy (i.e., 
$\sex\to\sex^*$)
and heavy-to-light (i.e., $\sex\to\ten$) baryon transitions.

   In HQET the $\half^+\to {3\over 2}^+$ matrix elements are given by
(see e.g., \cite{Yan})
\be
\la B_f^*(v')|V_\mu|B_i(v)\ra &=& {1\over\sqrt{3}}\,\bar u_f^\nu(v')\Big\{
(2g_{\mu\nu}+\gamma_\mu v_\nu)\xi_1   
+ v_\nu[(1-v\cdot v')\gamma_\mu-2v'_\mu]\xi_2\Big\}\gamma_5 u_i(v),   \non \\
\la B_f^*(v')|A_\mu|B_i(v)\ra &=& -{1\over\sqrt{3}}\,\bar u_f^\nu(v')\Big\{
(2g_{\mu\nu}-\gamma_\mu v_\nu)\xi_1   
+ v_\nu[(1+v\cdot v')\gamma_\mu-2v'_\mu]\xi_2\Big\}u_i(v),   \non \\
\en
where $\xi_1$ and $\xi_2$ are the baryon Isgur-Wise functions introduced
in (2.12). We find that at zero recoil 
\be
&& \bar f_1(q^2_m)={2\over\sqrt{3}},~~~\bar f_2(q^2_m)={1\over\sqrt{3}m_i},
~~~\bar f_3(q^2_m)=-{2\over\sqrt{3}}\,{\xi_2(1)\over m_im_f},   \non \\
&& \bar g_1(q^2_m)=-{2\over\sqrt{3}},~~~\bar g_2(q^2_m)={1\over\sqrt{3}m_i}[1-
2\xi_2(1)],~~~\bar g_3(q^2_m)={2\over\sqrt{3}}\,{\xi_2(1)\over m_im_f}.
\en
Since $N_{fi}=1$ for heavy-to-heavy transition, it is clear that the
quark model results for $\half^+\to {3\over 2}^+$ form factors (2.21) 
in the heavy quark limit are in agreement with the HQET predictions (2.23)
with $\xi_2(1)=\half$ [see Eq.~(2.16)].


    Since the calculation for the $q^2$ dependence of form factors is 
beyond the scope of the nonrelativistic quark model, we will follow the
conventional practice to assume a pole
dominance for the form-factor $q^2$ behavior:
\be
f(q^2)={f(0)\over \left(1-{q^2\over m_V^2}\right)^n}\,,~~~~g(q^2)={g(0)\over 
\left(1-{q^2\over m_A^2}\right)^n}\,,
\en
where $m_V$ ($m_A$) is the pole mass of the vector (axial-vector) meson 
with the same quantum number as the current under consideration.
The function 
\be
G(q^2)=\left({1-q^2_m/m^2_{\rm pole}\over 1-q^2/m_{\rm pole}^2} \right)^2
\non
\en
plays the role of the baryon Isgur-Wise function $\zeta(\omega)$ for 
$\Lambda_Q\to
\Lambda_{Q'}$ transition, namely $G=1$ at $q^2=q^2_m$. The function
$\zeta(\omega)$ has been calculated in the literature in various different
models \cite{Jenkins,Sadzi,GuoK,Iva}. Using the pole masses $m_V=6.34$ GeV, 
$m_A=6.73$ GeV for $\Lambda_b\to\Lambda_c$ transition, it is found
in \cite{CT96} that $G(q^2)$ is consistent with $\zeta(\omega)$ only if 
$n=2$. Nevertheless, one should bear in mind that the $q^2$ behavior of form
factors is probably more complicated and it is likely that a simple
pole dominance only applies to a certain $q^2$ region.

    Assuming a dipole $q^2$ behavior for form factors, we have tabulated
in Table I the numerical values of $B_b(\half^+)\to\half^+$, $B_b(\half^+)
\to{3\over2}^+$ and $B_c(\half^+)\to {3\over 2}^+$ form factors at
$q^2=0$ calculated using (2.10) and (2.21).
Uses have been made of $|V_{cb}|=0.038$ \cite{Stone}, the constituent 
quark masses (light quark masses being taken from p.619 of PDG \cite{PDG})
\be
m_b=5\,{\rm GeV},~~~m_c=1.6\,{\rm GeV},~~m_s=510\,{\rm MeV},~~m_d=322\,{\rm
MeV},~~m_u=338\,{\rm MeV},
\en
the pole masses:
\be
b\to c: &&~~~~~m_V=6.34\,{\rm GeV},~~~m_A=6.73\,{\rm GeV},   \non \\
b\to s: &&~~~~~m_V=5.42\,{\rm GeV},~~~m_A=5.86\,{\rm GeV},   \non  \\
b\to d: &&~~~~~m_V=5.32\,{\rm GeV},~~~m_A=5.71\,{\rm GeV},   \\
c\to s: &&~~~~~m_V=2.11\,{\rm GeV},~~~m_A=2.54\,{\rm GeV},   \non \\
c\to u: &&~~~~~m_V=2.01\,{\rm GeV},~~~m_A=2.42\,{\rm GeV},   \non 
\en
and the bottom baryon masses:
\be
m_{\Lambda_b}=5.621\,{\rm GeV},~~~m_{\Xi_b}=5.80\,{\rm GeV},~~~m_{\Omega_b}
=6.04\,{\rm GeV}.
\en
Note that the CDF measurement \cite{CDF96} $m_{\Lambda_b}=5621\pm 4\pm 3$ MeV 
has better accuracy than the PDG value $5641\pm 50$ MeV \cite{PDG}; the 
combined value is $m_{\Lambda_b}=5621\pm 5$ MeV.

\section{Results and Discussion}
  With the baryon form factors tabulated in Table I we are in a position
to compute the factorizable contributions to the decay rate and up-down
asymmetry for Cabibbo-allowed weak decays of bottom baryons $B_b(\half^+)\to
\half^+({3\over 2}^+)+P(V)$. The factorizable external and internal
$W$-emission amplitudes are given by (2.6), (2.7), (2.18) and (2.19).
The calculated results are summarized in Tables II and III. (The formulas for
decay rates and up-down asymmetries are given in Appendix B.) For decay
constants we use
\be
&& f_\pi=132\,{\rm MeV},~~~~~f_D=200\,{\rm MeV}~\cite{Sach},~~~~~f_{D_s}=
241\,{\rm MeV}~\cite{Sach},   \non \\
&& f_\rho=216\,{\rm MeV},~~~~~f_{J/\psi}=395\,{\rm MeV},~~~~~f_{\psi'}=
293\,{\rm MeV},
\en
where we have taken into account the momentum dependence of the fine-structure
constant to determine $f_{J/\psi}$ and $f_{\psi'}$ from experiment. In the 
absence of reliable theoretical estimates for $f_{D^*}$ and $f_{D^*_s}$, we 
have taken $f_{D^*}=f_{D}$ and $f_{D^*_s}=f_{D_s}$ for numerical calculations. 


    From Tables II and III we see that, except for those decay modes with 
$\psi'$ in the final state and for $\Omega_b\to\half^++P(V)$ decays,
the up-down asymmetry parameter $\alpha$ is found to be negative.\footnote{The
parameter $\alpha$ of $\Lambda_b\to J/\psi\Lambda$ is estimated to be 0.25
in \cite{Datta}, whereas it is $-0.10$ in our case.}
As noted in \cite{Mannel}, the parameter $\alpha$ in 
$\half^+\to\half^++P(V)$ decay becomes
$-1$ in the soft pseudoscalar meson or vector meson limit, i.e., $m_P\to
0$ or $m_V\to 0$. In practice, $\alpha$ is sensitive to $m_V$ but not so
to $m_P$. For example, $\alpha\approx -1$ for $\Lambda_b\to D_s\Lambda_c$
and $\Xi_b\to D_s\Xi_c$ even though the $D_s$ meson is heavy, but it changes
from $\alpha=-0.88$ for $\Lambda_b\to\rho\Lambda_c$ to $-0.10$ for $\Lambda_b
\to J/\psi\Lambda$. As stressed in Sec.~II, by treating $a_1$ and $a_2$ 
as free parameters, our predictions should be most reliable
for those decay modes which proceed only through the external $W$-emission
diagram $\a$ or the internal $W$-emission $\b'$.
Moreover, we have argued that the penguin contributions $\e'$ and $\e$ to
Cabibbo-allowed decays are safely negligible and that
the $W$-exchange amplitudes $\co,~\ct,~\cp$
are very likely to be suppressed in bottom baryon decays. It is thus very
interesting to test the suppression of $W$-exchange in decay modes of 
$B_b(\three)\to B(\ten)+P(V)$ that proceed only through $W$-exchange
[see (2.3)] and in decays $B_b(\three)\to B_c(\three)+P(V)$, e.g.,
$\Xi_b\to\pi(\rho)\Xi_c,~\Xi_b\to D_s^{(*)}\Xi_c$, that receive 
contributions from factorizable terms and $W$-exchange. Since the
nonfactorizable internal $W$-emission amplitude $\b$ is {\it a priori} not
negligible, our results for $\Lambda_b\to\pi(\rho)\Lambda_c,~\Omega_b\to
D_s^{(*)}\Omega_c^{(*)}$ (see Tables II and III) are subject to the
uncertainties due to possible contributions from the quark diagram $\b$.

   In order to have the idea about the
magnitude of branching ratios, let us take $a_1\sim 1$ as that inferred from
$B\to D^{(*)}\pi(\rho)$ decays \cite{CT95} and $a_2\sim 0.28$ as that in $B\to
J/\psi K^{(*)}$ decays.\footnote{A fit to recent measurements of $B\to 
J/\psi K(K^*)$ decays by CDF and CLEO yields \cite{Cheng96} $a_2(B\to 
J/\psi K)=0.30$ and $a_2(B\to J/\psi K^*)=0.26$.}
Using the current world average $\tau(\Lambda_b)=(1.23\pm 0.06)\times 
10^{-12}s$ \cite{Stone}, we find from Table II that
\be
&& \b(\Lambda_b^0\to D_s^-\Lambda_c^+) \cong 1.1\times 10^{-2},~~~~~\b(
\Lambda_b^0\to D_s^{*-}\Lambda_c^+) \cong 9.1\times 10^{-3},   \non \\
&& \b(\Lambda_b^0\to\pi^-\Lambda_c^+)\sim 3.8\times 10^{-3},~~~~~
\b(\Lambda_b^0\to\rho^-\Lambda_c^+) \sim 5.4\times 10^{-3},   \\
&& \b(\Lambda_b^0\to J/\psi \Lambda)=1.6\times 10^{-4},~~~~~\b(\Lambda_b^0\to 
\psi'\Lambda)=1.4\times 10^{-4}.   \non 
\en
Our estimate for the branching ratio of $\Lambda_b\to J/\psi \Lambda$
is consistent with the CDF result \cite{CDF96}:
\be
\b(\Lambda_b\to J/\psi \Lambda)=\,(3.7\pm 1.7\pm 0.4)\times 10^{-4}.
\en
Recall that the predictions (3.2) are obtained for $|V_{cb}|=0.038\,$.

   Since the decay mode $\Omega_c^0\to\pi^+\Omega^-$ has been seen 
experimentally, we also show the estimate of $\Gamma$ and $\alpha$ in 
Table IV for $\Omega_c^0\to {3\over 2}^++P(V)$ decays with the relevant 
form factors being given in Table I.  For comparison, we have
displayed in Table IV the model results of Xu and Kamal 
\cite{XK92b}\footnote{The $B$ and $D$ amplitudes in Eq.~(4) of \cite{XK92b},
where the formulas for $\Gamma$ and $\alpha$ in $\half^+\to{3\over 2}^++P$ 
decay are given, should be interchanged.},
K\"orner and Kr\"amer \cite{Korner}. In the model of Xu and Kamal, 
the $D$-wave amplitude in (2.17) and hence the parameter $\alpha$
vanishes in the decay $\Omega_c\to {3\over 2}^++P$ due to the fact that
the vector current is conserved at all $q^2$ in their scheme 1 and at $q^2=0$
in scheme 2. By contrast, the $D$-wave amplitude in our case does not
vanish. Assuming that the form factors $\bar f_1,~\bar f_2,~\bar f_3$ 
have the same $q^2$ dependence, we see from (2.18) and (2.21) that 
the amplitude $D$ is proportional to $(E_f-m_f)/m_f$, which vanishes at
$q^2=q^2_{\rm max}$ but not at $q^2=m_P^2$. Contrary to the decay 
$\Omega_b^-\to {3\over 2}^++P(V)$, the
up-down asymmetry is found to be positive in $\Omega_c^0\to{3\over 2}^+
+P(V)$ decays. Note that the sign of $\alpha$ for $\Omega_c\to 
{3\over 2}^++V$ is 
opposite to that of \cite{XK92b}.\footnote{It seems to us that the sign 
of $A_i$ and $B_i$ (or $C_i$ and $D_i$ in our notation) in
Eq.~(58) of \cite{XK92b} should be flipped. A consequence of this sign
change will render $\alpha$ positive in $\Omega_c\to {3\over 2}^++V$ decay.}
Therefore, it is desirable to measure the parameter $\alpha$ in 
decays $\Omega_c\to{3\over 2}^++P(V)$ to discern 
different models. To have an estimate of the branching ratio, we take
the large-$N_c$ values $a_1(m_c)=1.10,~a_2(m_c)=-0.50$ as an illustration
and obtain
\be
&& {\cal B}(\Omega_c^0\to \pi^+\Omega^-)\simeq 1.0\times 10^{-2},~~~~~
{\cal B}(\Omega_c^0\to \rho^+\Omega^-)\simeq 3.6\times 10^{-2},   \non \\
&& {\cal B}(\Omega_c^0\to \overline K^0\Xi^{*0})\simeq 2.5\times 10^{-3},~~~~~
{\cal B}(\Omega_c^0\to \overline K^{*0}\Xi^{*0})\simeq 3.7\times 10^{-3},
\en
where use of $\tau(\Omega_c)=6.4\times 10^{-14}s$ \cite{PDG} has been made.


   Three important ingredients on which the calculations are built in this 
work are : factorization, nonrelativistic quark model, and diople $q^2$
behavior of form factors. The factorization hypothesis can be tested by
extracting the effective parameters $a_1$, $a_2$ from data and 
seeing if they are channel independent. Thus far we have neglected the
effects of final-state interactions which are supposed to be less important
in bottom baryon decay since decay particles in the two-body final state
are energetic and moving fast,
allowing less time for significant final-state interactions. We have argued 
that, in the nonrelativistic quark model, the ratio of (k.e./$M_q$) is small
even for the constituent quark inside the nucleon (or hyperon) at rest.
As for the $q^2$ dependence of baryon form factors, we have applied
dipole dominance motivated by the consistency with the $q^2$ behavior
of the baryon Isgur-Wise function. Nevertheless, in order to check the
sensitivity of the form factor $q^2$ dependence, we have repeated 
calculations using the monopole form. Since for a given $q^2$, the absolute
values of the form factors in the monopole behavior are larger than that in 
the dipole one, it is expected that the branching rations obtained
under the monopole ansatz will get enhanced, especially when the final-state
baryons are hyperons. Numerically, we find that, while decay asymmetries
remain essentially unchanged, the decay rates of $B_b({1\over 2}^+)\to
B_c({1\over 2}^+)+P(V)$ and $B_b({1\over 2}^+)\to {\rm hyperon}+P(V)$ are in 
general enhanced by factors of $\sim 1.8$ and $\sim 3.5$, respectively.
In reality, the utilization of a simple $q^2$
dependence, monopole or dipole, is probably too simplified. It thus
appears that major calculational uncertainties arise mainly from the {\it 
ad hoc} ansatz on the form factor $q^2$ behavior.

   In conclusion, if the $W$-exchange contribution to the hadronic decays of
bottom baryons is negligible, as we have argued, then the theoretical
description of bottom baryons decaying into $\half^++P(V)$ and
${3\over 2}^++P(V)$ is relatively clean since these decays either receive 
contributions only from external/internal $W$-emission or 
are dominated by factorizable terms. The absence or the suppression of 
the so-called pole terms makes the study of Cabibbo-allowed decays of bottom
baryons considerably simpler than that in charmed baryon decay. We have 
evaluated the heavy-to-heavy and heavy-to-light baryon form factors at
zero recoil using the nonrelativistic quark model and reproduced the HQET
results for heavy-to-heavy baryon transition. It is stressed that for
heavy-to-light baryon form factors, there is a flavor-suppression factor 
which must be taken into account in calculations. Predictions of the 
decay rates and up-down asymmetries for $B_b\to\half^++P(V)$ 
and $\Omega_c\to{3\over 2}^++P(V)$ are given. The parameter 
$\alpha$ is found to be negative except for $\Omega_b\to\half^++P(V)$ decays 
and for those decay modes with $\psi'$ in the final state. We also
present estimates of $\Gamma$ and $\alpha$ for $\Omega_c\to{3\over 2}^++P(V)$
decays. It is very desirable to measure the asymmetry parameter to discern
different models.

\vskip 1.5 cm
\centerline{\bf ACKNOWLEDGMENT}
\vskip 0.5cm
    This work was supported in part by the National Science Council of ROC
under Contract No. NSC86-2112-M-001-020.

\vskip 1.0 cm
\centerline{\bf Appendix A.~~Baryon Form Factors in the Quark Model}
\vskip 0.5 cm
\renewcommand{\thesection}{\Alph{section}}
\renewcommand{\theequation}{\thesection\arabic{equation}}
\setcounter{equation}{0}
\setcounter{section}{1}
   Since the $\half^+$ to $\half^+$ baryon form factors have been evaluated
at zero recoil in the nonrelativistic quark model \cite{CT96}, we will focus
in this Appendix on the baryon form factors in $\half^+$ to ${3\over 2}^+$ 
transition. Let $u^\alpha$ be the Rarita-Schwinger vector-spinor for a
spin-${3\over 2}$ particle. The general four plane-wave
solutions for $u^\alpha$ are (see, for example, \cite{Lurie})
\be
&& u^\alpha_1=(u^0_1,\,\vec{u}_1)=(0,\,\vec{\epsilon}_1u_\up),   \non \\
&& u^\alpha_2=(u^0_2,\,\vec{u}_2)=\left(\sqrt{2\over 3}\,{|p|\over m}u_\up,\,
{1\over\sqrt{3}}\vec{\epsilon}_1u_\dw-\sqrt{2\over 3}\,{E\over m}\vec{
\epsilon}_3u_\up\right),   \non \\
&& u^\alpha_3=(u^0_3,\,\vec{u}_3)=\left(\sqrt{2\over 3}\,{|p|\over m}u_\dw,\,
{1\over\sqrt{3}}\vec{\epsilon}_2u_\up-\sqrt{2\over 3}\,{E\over m}\vec{
\epsilon}_3u_\dw\right),   \\
&& u^\alpha_4=(u^0_4,\,\vec{u}_4)=(0,\,\vec{\epsilon}_2u_\dw),   \non 
\en
in the frame where the baryon momentum $\vec{p}$ is along the $z$-axis, and
\be
\epsilon_1={1\over\sqrt{2}}\left(\matrix{ 1 \cr i  \cr 0 \cr}\right),~~~
\epsilon_2={1\over\sqrt{2}}\left(\matrix{ 1 \cr -i  \cr 0 \cr}\right),~~~
\epsilon_3=\left(\matrix{ 0 \cr 0  \cr 1 \cr}\right).
\en
Note that the spin $z$-component of the four solutions (A1) corresponds to
${3\over 2},\half,-\half,-{3\over 2}$, respectively. Substituting (A1)
into (2.20) yields
\be
\la B^*_f(+1/2)|V_0|B_i(+1/2)\ra &=& \sqrt{2\over 3}\,{p\over m_f}\bar{u}
_\up(\bar f_1\gamma_5+\bar f_2m_i\gamma_0\gamma_5+\bar f_3m_iE_f\gamma_5)
u_\up,   \\
\la B^*_f(+1/2)|A_0|B_i(+1/2)\ra &=& \sqrt{2\over 3}\,{p\over m_f}\bar{u}
_\up(\bar g_1+\bar g_2m_i\gamma_0+\bar g_3m_iE_f)u_\up,   \\
\la B^*_f(+3/2)|\vec{V}|B_i(+1/2)\ra &=& -\bar f_1\vec{\epsilon}_1\bar{u}_\up
\gamma_5u_\up,  \\
\la B^*_f(+3/2)|\vec{A}|B_i(+1/2)\ra &=& -\bar g_1\vec{\epsilon}_1\bar{u}_\up 
u_\up,    \\
\la B^*_f(+1/2)|\vec{V}|B_i(-1/2)\ra &=& -\bar f_1\left({1\over\sqrt{3}}\vec{
\epsilon}_1\bar{u}_\dw-\sqrt{2\over 3}\,{E_f\over m_f}\vec{\epsilon}_3\bar{u}
_\up\right)\gamma_5u_\dw   \non \\
&& +\sqrt{2\over 3}\,{pm_i\over m_f}\,\bar{u}_\up(\bar f_2\vec{\gamma}
\gamma_5+\bar f_3\vec{p}\gamma_5)u_\dw,    \\
\la B^*_f(+1/2)|\vec{A}|B_i(-1/2)\ra &=& -\bar g_1\left({1\over\sqrt{3}}\vec{
\epsilon}_1\bar{u}_\dw-\sqrt{2\over 3}\,{E_f\over m_f}\vec{\epsilon}_3\bar{u}
_\up\right)u_\dw    \non \\
&& +\sqrt{2\over 3}\,{pm_i\over m_f}\,\bar{u}_\up(\bar g_2\vec{\gamma}+\bar 
g_3\vec{p}\,)u_\dw,
\en
where $\vec{p}$ is the momentum of the daughter baryon along the $z$-axis in 
the rest frame of the parent baryon. The baryon matrix elements in 
(A3)-(A8) can be evaluated in the nonrelativistic quark model. Following 
the same procedure outlined in \cite{CT96}, we obtain
\be
\la B_f^*|V_0|B_i\ra/N_f &=& \la 1\ra,   \non \\
\la B_f^*|\vec{V}|B_i\ra/N_f &=& -{1\over 2m_q}\left(1-{\bar{\Lambda}\over 2
m_f}\right)\la\vec{q}+i\vec{\sigma}\times\vec{q}\,\ra+{\bar{\Lambda}\over 4m_Q
m_f}\la\vec{q}-i\vec{\sigma}\times\vec{q}\,\ra,   \non \\
\la B_f^*|A_0|B_i\ra/N_f &=& \left[-{1\over 2m_q}\left(1-{\bar{\Lambda}\over 2
m_f}\right)+{\bar{\Lambda}\over 4m_Qm_f}\right]\la\vec{\sigma}\cdot\vec{q}\,
\ra,  \\
\la B_f^*|\vec{A}|B_i\ra/N_f &=& \la\vec{\sigma}\ra-{\bar{\Lambda}\over 4m_Q
m_f^2}\,\la(\vec{\sigma}\cdot\vec{q}\,)\vec{q}-{1\over 2}\vec{\sigma}q^2\ra,   
\non
\en
where $\vec{q}=\vec{p}_i-\vec{p}_f$, $N_f=\sqrt{(E_f+m_f)/2m_f}$,
$m_q$ is the mass of the quark $q$ in $B_f^*$ coming from the decay
of the heavy quark $Q$ in $B_i$, and $\la X\ra$ stands for $_{\rm 
flavor-spin}\la B^*_f|X|B_i\ra_{\rm flavor-spin}$. Form factors $\bar f_i$ 
and $\bar g_i$ are then determined from (A3) to (A9). For example, $\bar
f_1$ can be determined from the $x$ (or $y$) component of (A5) which is 
\be
\la B_f^*(+3/2)|V_x|B_i(+1/2)\ra=-{\bar f_1\over\sqrt{2}}\,{N_f\over 
E_f+m_f}\,\chi^\dagger_\up\vec{\sigma}\cdot\vec{q}\chi_\up={\bar f_1\over
\sqrt{2}}\,{pN_f\over E_f+m_f},
\en
where $\chi$ is a two-component Pauli spinor. From (A9) we find
\be
 \la B_f^*(+3/2)|V_x|B_i(+1/2)\ra = {pN_f\over 4m_q}\left(1-{\bar{\Lambda}
\over 2m_f}+{\bar\Lambda m_q\over 2m_Qm_f}\right)\la(\sigma_+-\sigma_-)
b_q^\dagger b_Q\ra.
\en
Since [$N_{fi}$ being defined by (2.11)]
\be
_{\rm flavor-spin}\la B_f^*(+3/2)|(\sigma_+-\sigma_-)b_q^\dagger b_Q|B_i(+1/2)
\ra_{\rm flavor-spin}={4\over\sqrt{6}}N_{fi},
\en
for sextet $B_i$ and vanishes for antitriplet $B_i$, it is evident that
only the decay of $\Omega_b$ into ${3\over 2}^++P(V)$ can receive
factorizable contributions. Indeed the decays $B_b(\three)\to B(\ten)+P(V)$
proceed only through $W$-exchange or $W$-loop, as discussed in Sec.~II.
It follows from (A10)-(A12) that at zero recoil
\be
\bar f_1(q_m^2)/N_{fi}=\,{2\over\sqrt{3}}\left(1+{\bar{\Lambda}\over 
2m_q}+{\bar{\Lambda}\over 2m_Q}\right),
\en
which is the result shown in (2.21). The form factor $\bar f_2$ is then fixed
by the $x$ (or $y$) component of (A7). Substituting $\bar f_1$ and $\bar f_2$ 
into (A3) determines $\bar f_3$. The remaining form factors $\bar g_i$ are
determined in a similar way.

\vskip 1.0 cm
\centerline{\bf Appendix B.~~Kinematics}
\vskip 0.7 cm
\setcounter{equation}{0}
\setcounter{section}{2}
 In this Appendix we summarize the kinematics relevant to the two-body
hadronic decays of $\half^+\to\half^+({3\over 2}^+)+P(V)$.
   With the amplitudes (2.4) for $\half^+\to\half^++P$ decay and (2.17)
for $\half^+\to {3\over 2}^++P$, the decay rates and up-down asymmetries read
\be
\Gamma (1/2^+\to 1/2^++P) &=& {p_c\over 8\pi}\left\{ {(m_i+m_f)^2-m_P^2
\over m_i^2}\,|A|^2+{(m_i-m_f)^2-m^2_P\over m_i^2}\,|B|^2\right\},  \non \\
\alpha (1/2^+\to 1/2^++P) &=& -{2\kappa{\rm Re}(A^*B)\over |A|^2+
\kappa^2|B|^2},
\en
and
\be
\Gamma (1/2^+\to {3/ 2}^++P) &=& {p_c^3\over 8\pi}\left\{ {(m_i-m_f)^2-
m_P^2\over m_i^2}\,|C|^2+{(m_i+m_f)^2-m^2_P\over m_i^2}\,|D|^2\right\},\non \\
\alpha (1/2^+\to{3/ 2}^++P) &=& -{2\kappa{\rm Re}(C^*D)\over \kappa^2
|C|^2+|D|^2},
\en
where $p_c$ is the c.m. momentum and $\kappa=p_c/(E_f+m_f)=\sqrt{(E_f-m_f)/(
E_f+m_f)}$. For $\half^+\to\half^++V$ decay we have \cite{Pak}
\footnote{The formulas for the decay rate of $\half^+\to\half^++V$ decay given
in \cite{CT92,CT96} contain some errors which are corrected in errata.}
\be
\Gamma(1/2^+\to 1/2^++V) &=& {p_c\over 8\pi}\,{E_f+m_f\over m_i}\left[
2(|S|^2+|P_2|^2)+{E^2_V\over m_V^2}(|S+D|^2+|P_1|^2)\right], \non \\
\alpha(1/2^+\to 1/2^++V) &=& {4m^2_V{\rm Re}(S^*P_2)+2E^2_V{\rm Re}(S+D)^*
P_1\over 2m_V^2(|S|^2+|P_2|^2)+E^2_V(|S+D|^2+|P_1|^2)},
\en
with the $S,~P$ and $D$ waves given by
\be
S &=& -A_1,   \non \\
P_1 &=& -{p_c\over E_V}\left( {m_i+m_f\over E_f+m_f}B_1+m_iB_2\right), \non \\
P_2 &=& {p_c\over E_f+m_f}B_1,   \\
D &=& -{p_c^2\over E_V(E_f+m_f)}\,(A_1-m_iA_2), \non
\en
where the amplitudes $A_1,~A_2,~B_1$ and $B_2$ are defined in (2.4).
However, as emphasized in \cite{Korner}, it is also convenient to express 
$\Gamma$ and $\alpha$ in terms of the helicity amplitudes
\be
h_{\lambda_f,\lambda_V;\lambda_i}=\,\la B_f(\lambda_f)V(\lambda_V)|H_W|B_i(
\lambda_i)\ra
\en
with $\lambda_i=\lambda_f-\lambda_V$. Then \cite{Korner}
\be
\Gamma &=& {p_c\over 32\pi m_i^2}\sum_{\lambda_f,\lambda_V}\left(|h_{
\lambda_f,\lambda_V;1/2}|^2-|h_{-\lambda_f,-\lambda_V;-1/2}|^2\right), \non\\
\alpha &=& \sum_{\lambda_f,\lambda_V} { \left(|h_{
\lambda_f,\lambda_V;1/2}|^2-|h_{-\lambda_f,-\lambda_V;-1/2}|^2\right)\over
\left(|h_{\lambda_f,\lambda_V;1/2}|^2+|h_{-\lambda_f,-\lambda_V;-1/2}|^2
\right)}\,.
\en
The helicity amplitudes for $\half^+\to\half^++V$ decay are given by 
\cite{Korner}
\be
H_{\lambda_1,\lambda_2;1/2}^{ {\rm p.v.}\,({\rm p.c.})}
&=& H_{\lambda_1,\lambda_2;1/2}\mp H_{-\lambda_1,-\lambda_2;-1/2},   \non\\
H_{-1/2,-1;1/2}^{ {\rm p.v.}\,({\rm p.c.}) } &=& 
2\left\{ \matrix{\sqrt{Q_+}A_1  \cr -\sqrt{Q_-}B_1  \cr}\right\},   \\
H_{1/2,0;1/2}^{ {\rm p.v.}\,({\rm p.c.}) } &=& 
{\sqrt{2}\over m_V}\left\{ \matrix{\sqrt{Q_+}\,(m_i-m_f)A_1-\sqrt{Q_-}\,m_ip_c
A_2\cr  \sqrt{Q_-}\,(m_i+m_f)B_1+\sqrt{Q_+}\,m_ip_cB_2  \cr} \right\},   \non
\en
where the upper (lower) entry is for parity-violating (-conserving) helicity
amplitude, and
\be
Q_\pm=\,(m_i\pm m_f)^2-m^2_V=2m_i(E_f\pm m_f).
\en
Note that the helicity amplitudes for $\half^+\to \half^++V$ decay shown in 
Eq.~(20) of \cite{Korner} are too large by a factor of $\sqrt{2}$.
One can check explicitly that the decay rates and up-down asymmetries 
evaluated 
using the partial-wave method and the helicity-amplitude method are 
equivalent. For completeness, we also list the helicity amplitudes for
$\half^+\to{3\over 2}^++V$ decay \cite{Korner}:
\be
H_{\lambda_1,\lambda_2;1/2}^{ {\rm p.v.}\,({\rm p.c.})}
&=& H_{\lambda_1,\lambda_2;1/2}\pm H_{-\lambda_1,-\lambda_2;-1/2},   \non \\
H_{3/2,1;1/2}^{ {\rm p.v.}\,({\rm p.c.}) } &=& 
2\left\{ \matrix{-\sqrt{Q_+}C_1  \cr \sqrt{Q_-}D_1  \cr} \right\},   \\
H_{-1/2,-1;1/2}^{ {\rm p.v.}\,({\rm p.c.}) } &=& 
{2\over \sqrt{3}}\left\{ \matrix{-\sqrt{Q_+}\,[C_1-2(Q_-/m_f)C_2]   \cr
\sqrt{Q_-}\,[D_1-2(Q_+/m_f)D_2]   \cr} \right\},   \non  \\
H_{1/2,0;1/2}^{ {\rm p.v.}\, ({\rm p.c.}) } &=& 
{2\sqrt{2}\over \sqrt{3}\,m_fm_V}\left\{ \matrix{-\sqrt{Q_+}\,\big[\half(m_i^2
-m_f^2-m_V^2)C_1 +Q_-(m_i+m_f)C_2+m_i^2p_c^2C_3\big]   \cr
\sqrt{Q_-}\,\big[\half(m_i^2-m_f^2-m_V^2)D_1  -Q_+(m_i-m_f)D_2+
m_i^2p_c^2D_3\big]   \cr} \right\}.   \non
\en

\renewcommand{\baselinestretch}{1.1}
\newcommand{\bi}{\bibitem}

\newpage

\vskip 0.5cm
\begin{table}[t]
{\small Table I.~~Nonrelativistic quark model predictions for
baryonic form factors evaluated at $q^2=0$ using dipole $q^2$ dependence 
($m_i$ being the mass of the parent
heavy baryon). Also shown are the spin and flavor factors
for various baryonic transitions.$^*$}
\begin{center}
\begin{tabular}{|l||c c c r c r r r|} \hline
Transition & $\eta$ & $N_{fi}$ & $f_1(0)$ & $f_2(0)m_i$ & $f_3(0)m_i$ & 
$g_1(0)$ & $g_2(0)m_i$ & $g_3(0)m_i$   \\  \hline\hline
$\Lambda_b^0\to\Lambda_c^+$ & 1 & 1 & 0.530 & $-0.100$ & $-0.012$ & 0.577 &
$-0.013$ & $-0.109$ \\
$\Lambda_b^0\to\Lambda^0$ & 1 & ${1\over\sqrt{3}}$ & 0.062 & $-0.025$ & 
$-0.008$ & 0.108 & $-0.014$ & $-0.043$ \\
$\Lambda_b^0\to n$ & 1 & ${1\over\sqrt{2}}$ & 0.045 & $-0.024$ & $-0.011$ & 
0.095 & $-0.022$ & $-0.051$ \\  \hline
$\Xi_b^{0,-}\to\Xi_c^{+,0}$ & 1 & 1 & 0.533 & $-0.124$ & $-0.018$ & 0.580 &
$-0.019$ & $-0.135$ \\
$\Xi_b^{0,-}\to\Xi^{0,-}$ & 1 & ${1\over\sqrt{2}}$ & 0.083 & $-0.041$ & 
$-0.016$ & 0.143 & $-0.027$ & $-0.070$ \\
$\Xi_b^{0,-}\to\Sigma^{0,-}$ & 1 & ${1\over 2}$ & 0.042 & $-0.028$ & $-0.014$ 
& 0.083 & $-0.028$ & $-0.054$ \\  
$\Xi_b^{0}\to\Lambda^{0}$ & 1 & ${1\over 2\sqrt{3}}$ & 0.019 & $-0.012$ & 
$-0.006$ & 0.041 & $-0.013$ & $-0.025$ \\ \hline  
$\Omega_b^-\to \Omega_c^0$ & $-{1\over 3}$ & 1 & 0.710 & 0.666 & $-0.339$ &
$-0.195$ & 0.009 & 0.056   \\
$\Omega_b^-\to \Xi^-$ & $-{1\over 3}$ & ${1\over\sqrt{3}}$ & 0.102 & 0.103 & 
$-0.097$ & $-0.028$ & 0.011 & 0.019   \\ \hline
$\Omega_b^-\to \Omega_c^{*0}$ &  & 1 & 0.902 & 0.451 & $-0.451$ &
$-0.606$ & $-0.237$ & 0.490   \\
$\Omega_b^-\to \Omega^-$ &  & 1 & 0.320 & 0.160 & $-0.160$ &
$-0.228$ & $-0.260$ & 0.257   \\
$\Omega_b^-\to \Xi^{*-}$ &  & ${1\over\sqrt{3}}$ & 0.158 & 0.079 & $-0.079$ &
$-0.094$ & $-0.177$ & 0.141   \\  \hline
$\Omega_c^0\to \Omega^-$ & & 1 & 1.167 & 0.837 & $-0.837$ & $-0.804$ &
$-0.916$ & 1.006  \\
$\Omega_c^0\to \Xi^{*0}$ & & ${1\over\sqrt{3}}$ & 0.942 & 0.471 & $-0.471$ & 
$-0.390$ & $-0.731$ & 0.634  \\  \hline
\end{tabular}
\end{center}
$^*$ {\small Our flavor factors
$N_{fi}$ for $\Omega_c^0\to\Omega^-$ and $\Omega_c^0\to\Xi^{*0}$ are two
times smaller than that in \cite{HK}.} 
\end{table}

\begin{table}
{\small Table II.~~Factorizable contributions to the decay rates (in units of 
$10^{10}s^{-1}$) and up-down asymmetries of Cabibbo-allowed 
nonleptonic weak decays of bottom baryons ${1\over 2}^+\to {1\over 2}^++P(V)$.
Also shown are the quark-diagram amplitudes for various reactions.}
\begin{center}
\begin{tabular}{|l c r r|l c r r|} \hline
Decay & Diagram & $\Gamma$ & $\alpha$ & Decay & Diagram & 
$\Gamma$ & $\alpha$  \\  \hline\hline
$\Lambda_b^0\to\pi^-\Lambda_c^+$ & $\a,\b,\co,\ct$ & $0.31a_1^2$ &
$-0.99$ & $\Xi_b^{0,-}\to\pi^-\Xi_c^{+,0}$ & $\a,\ct$ & $0.33a_1^2$ & 
$-1.00$ \\
$\Lambda_b^0\to\rho^-\Lambda_c^+$ & $\a,\b,\co,\ct$ & $0.44a_1^2$ &
$-0.88$ & $\Xi_b^{0,-}\to\rho^-\Xi_c^{+,0}$ & $\a,\ct$ & $0.47a_1^2$ & $-0.88$ 
\\
$\Lambda_b^0\to D_s^-\Lambda_c^+$ & $\a,\e'$ & $0.93a_1^2$ &
$-0.99$ & $\Xi_b^{0,-}\to D_s^-\Xi_c^{+,0}$ & $\a,\cp,\e',\e$ & $0.99a_1^2$ & 
$-0.99$ \\
$\Lambda_b^0\to D_s^{*-}\Lambda_c^+$ & $\a,\e'$ & $0.74a_1^2$ &
$-0.36$ & $\Xi_b^{0,-}\to D_s^{*-}\Xi_c^{+,0}$ & $\a,\cp,\e',\e$ 
& $0.78a_1^2$ & $-0.36$ \\ 
$\Lambda_b^0\to J/\psi\Lambda^0$ & $\bp$ & $0.17a_2^2$ &
$-0.10$ & $\Xi_b^{0,-}\to J/\psi\Xi^{0,-}$ & $\bp$ & $0.32a_2^2$ & 
$-0.10$ \\ 
$\Lambda_b^0\to \psi'\Lambda^0$ & $\bp$ & $0.14a_2^2$ &
$0.05$ & $\Xi_b^{0,-}\to \psi'\Xi^{0,-}$ & $\b'$ & $0.27a_2^2$ & 
$0.05$ \\ 
$\Lambda_b^0\to D^0 n$ & $\bp,\c'$ & $0.024a_2^2$ &
$-0.81$ & $\Xi_b^{0,-}\to D^0\Sigma^{0,-}$ & $\bp(\cp)^\dagger$ & $0.020a_2^2$ 
& $-0.85$ \\ 
$\Lambda_b^0\to D^{*0}n$ & $\bp,\cp$ & $0.017a_2^2$ &
$-0.42$ & $\Xi_b^{0,-}\to D^{*0}\Sigma^{0,-}$ & $\bp(\cp)^\dagger$ & 
$0.014a_2^2$ & $-0.45$ \\ 
$\Xi_b^0\to D^0\Lambda^0$ & $\bp$ & $0.005a_2^2$ & $-0.81$ &
$\Xi_b^0\to D^{*0}\Lambda^0$ & $\bp$ & $0.003a_2^2$ & $-0.44$ \\ \hline
$\Omega_b^-\to\pi^-\Omega_c^0$ & $\a$ & $0.30a_1^2$ & 0.51 & $\Omega_b^-\to
D_s^{*-}\Omega_c^0$ & $\a,\b,\e',\e$ & $0.35a_1^2$ & 0.64 \\
$\Omega_b^-\to\rho^-\Omega_c^0$ & $\a$ & $0.39a_1^2$ & 0.53 & $\Omega_b^-\to
D^0\Xi^-$ & $\bp$ & $0.033a_2^2$ & 0.47 \\
$\Omega_b^-\to D_s^-\Omega_c^0$ & $\a,\b,\e',\e$ & $1.09a_1^2$ & 0.42 & 
$\Omega_b^-\to D^{*0}\Xi^-$ & $\bp$ & $0.014a_2^2$ & 0.54 \\
\hline
\end{tabular}
\end{center}
{\small $^\dagger$ The decay modes $\Xi_b^0\to D^0\Sigma^0,~D^{*0}
\Sigma^0$ also receive $W$-exchange contribution $\c'$.}
\end{table}

\begin{table}
{\small Table III.~~Predicted decay rates (in units of $10^{10}s^{-1}$) and 
up-down asymmetries for Cabibbo-allowed nonleptonic weak decays of 
the bottom baryon $\Omega_b^-
\to {3\over 2}^++P(V)$. Also shown are the quark-diagram amplitudes for 
various reactions.}
\begin{center}
\begin{tabular}{|l c r r|l c r r|} \hline
 Decay & Diagram & $\Gamma$ & $\alpha$~~ & ~~Decay & Diagram & 
$\Gamma$ & $\alpha$  \\  \hline\hline
$\Omega_b^-\to\pi^-\Omega_c^{*0}$ & $\a$ & $0.67a_1^2$ & $-0.38$ & $\Omega_b^-
\to J/\psi\Omega^-$ & $\b'$ & $3.15a_2^2$ & $-0.18$  \\
$\Omega_b^-\to\rho^-\Omega_c^{*0}$ & $\a$ & $0.95a_1^2$ & $-0.75$ & 
$\Omega_b^-\to \psi'\Omega^-$ & $\b'$ & $1.94a_2^2$ & 0.004  \\
$\Omega_b^-\to D_s^-\Omega_c^{*0}$ & $\a,\b,\e',\e$ & $0.88a_1^2$ & $-0.22$ & 
$\Omega_b^-\to D^0\Xi^{*-}$ & $\b'$ & $0.23a_2^2$ & $-0.80$  \\
$\Omega_b^-\to D_s^{*-}\Omega_c^{*0}$ & $\a,\b,\e',\e$ & $0.98a_1^2$ & $-0.31$ 
& $\Omega_b^-\to D^{*0}\Xi^{*-}$ & $\b'$ & $0.27a_2^2$ & $-0.38$  \\
\hline
\end{tabular}
\end{center}
\end{table}

\begin{table}
{\small Table IV.~~Predicted decay rates (in units of $10^{11}s^{-1}$) and 
up-down asymmetries (in parentheses) for Cabibbo-allowed nonleptonic 
weak decays of the charmed baryon $\Omega_c^0\to {3\over 2}^++P(V)$ in various
models. The model calculations of Xu and Kamal are done in two different
schemes \cite{XK92b}.}
\begin{center}
\begin{tabular}{|l|c|l|l|c|} \hline
Decay & This work & \multicolumn{2}{c|} {Xu \& Kamal \cite{XK92b}} &
K\"orner \& Kr\"amer \cite{Korner} \\ \hline\hline
$\Omega_c^0\to\pi^+\Omega^-$ & $1.33a_1^2(0.17)$ & $2.13a_1^2(0)$ & 
$2.09a_1^2(0)$ & $0.50a_1^2$   \\
$\Omega_c^0\to\rho^+\Omega^-$ & $4.68a_1^2(0.43)$ & $11.6a_1^2(-0.08)$ & 
$11.3a_1^2(-0.21)$ & $2.93a_1^2$   \\
$\Omega_c^0\to\overline{K}^0\Xi^{*0}$ & $1.53a_2^2(0.35)$ & $1.00a_2^2(0)$ & 
$0.89a_2^2(0)$ & $0.58a_2^2$   \\
$\Omega_c^0\to\overline{K}^{*0}\Xi^{*0}$ & $2.32a_2^2(0.28)$ & 
$4.56a_2^2(-0.09)$ & $4.54a_2^2(-0.27)$ & $3.30a_2^2$   \\
\hline
\end{tabular}
\end{center}
\end{table}

\end{document}